\documentclass[traditabstract]{aa}
\usepackage{txfonts}
\usepackage{natbib}
\usepackage{array}
\usepackage{url}
\bibpunct{(}{)}{;}{a}{}{,}
\usepackage{graphicx}

\newcommand\hdf{HD~49933}
\newcommand\muhz{\ensuremath{\mu\mathrm{Hz}}}
\newcommand\chisq{\ensuremath{\chi^2}}
\newcommand\ea{et al.}
\newcommand{\hrd}{HR diagram}

\newcommand{\tauzero}{\ensuremath{\tau_0}}
\newcommand{\taubcz}{\ensuremath{\tau_\mathrm{BCZ}}}
\newcommand{\tauhiz}{\ensuremath{\tau_\mathrm{HIZ}}}
\newcommand{\phibcz}{\ensuremath{\phi_\mathrm{BCZ}}}
\newcommand{\phihiz}{\ensuremath{\phi_\mathrm{HIZ}}}
\newcommand{\abcz}{\ensuremath{A_\mathrm{BCZ}}}
\newcommand{\ahiz}{\ensuremath{A_\mathrm{HIZ}}}

\makeatletter

\newcommand{\Rmnum}[1]{\expandafter\@slowromancap\romannumeral #1@}
\makeatother

\def\myfigure#1#2#3#4{
	\begin{figure#4}
	\resizebox{\hsize}{!}{\includegraphics{#1}}
	\caption{#2 \label{#3}}
	\end{figure#4}
}

\begin{document}

\title{Seismic detection of acoustic sharp features in the CoRoT
target \hdf}

\author{ 
     Anwesh~Mazumdar\inst{1}
\and
     Eric~Michel\inst{2}
\and
     H.~M.~Antia\inst{3}
\and
     Sebastien~Deheuvels\inst{2,4}
}

\offprints{anwesh@tifr.res.in}

\institute{
Homi Bhabha Centre for Science Education, TIFR,
V.~N.~Purav Marg, Mankhurd, Mumbai 400088, India.
\and 
LESIA, Observatoire de Paris, UMR8109 CNRS, Universit\'e Pierre et Marie
Curie, Universit\'e Denis Diderot, Place J.\ Janssen, 92195 Meudon, France.
\and 
Tata Institute of Fundamental Research, Homi Bhabha Road, 
Mumbai 400005, India.
\and 
Astronomy Department, Yale University, P.O. Box 208101, New Haven, 
CT 065208101, USA.
}

\date{}

\authorrunning{Mazumdar et al.}
\titlerunning{Seismic detection of acoustic sharp features in HD49933}

\abstract
{ 
The technique of determining the acoustic location of layers of sharp
changes in the sound speed inside a star from the oscillatory signal in
its frequencies is applied on a solar-type star, the CoRoT target, \hdf.
We are able to determine the acoustic depth of the second helium
ionisation zone of \hdf\ to be $794_{-68}^{+55}$\,s.  The acoustic depth
of the base of the convective zone is found to be
$1855_{-412}^{+173}$\,s where the large error bars reflect the ambiguity
in the result, which is difficult to determine with present precision on
the frequencies because of the intrinsically weak nature of the signal.
The positions of both these layers are consistent with those in a
representative stellar model of \hdf.
}


\keywords{stars: oscillations -- stars: interiors}
\keywords{Stars: individual: HD~49933 -- Stars: oscillations
-- Stars: interiors -- Stars: fundamental parameters} 

\maketitle

\section{Introduction}
\label{sec:intro}

Any localised feature in the sound speed inside a star, such as that
caused by the change in the temperature gradient at the base of the
convection zone, introduces an oscillatory term in frequencies as a
function of radial order $n$, that is proportional to
$\sin(2\tau_\mathrm{d}\omega_{n,\ell} + \phi)$, \citep{gough90} where
$\tau_\mathrm{d}=\int^R_{r_\mathrm{d}} {dr\over c}$ is the acoustic
depth of the localised feature, $c$ is the speed of sound,
$r_\mathrm{d}$ is the radial distance where the feature is located,
$\omega_{n,\ell}$ is the angular frequency of a mode with radial order
$n$ and degree $\ell$, and $\phi$ is a phase factor.  This oscillatory
signature has been extensively studied for the Sun in order to determine
the extent of overshoot below the solar convection zone
\citep{mct94,ban94,rv94}. It has been proposed earlier that this may be
used for distant stars also to find the position of the base of the
convective envelope or the second helium ionisation zone
\citep{mct00,ma01,rv03,btg04,basu04,am05}.  Indeed, \citet{miglio10}
have used the modulation of the frequency separations to determine the
location of the second helium ionisation zone in a red giant star.

We present here the results of applying this technique for the first
time on a main sequence solar-type star: \hdf, observed by the CoRoT
mission in 2007--08 \citep{baglin06}.

\section{The technique}
\label{sec:technique}

The oscillatory signal in the frequencies is quite small and is embedded
in the frequencies together with a smooth trend arising from the regular
variation of the sound speed in the stellar interior.  It can be
enhanced by using the second differences $\delta^2 \nu(n,\ell) =
\nu(n-1,\ell) - 2\nu(n,\ell) + \nu(n+1,\ell)$, instead of the
frequencies themselves \citep[see, e.g.,][]{gough90,ban94,ma01,basu04}.

The acoustic depths of the base of the convective zone (BCZ) and the
second helium ionisation zone (HIZ), \taubcz\ and \tauhiz, respectively,
can be obtained from the data by fitting the second differences to a
suitable function representing the oscillatory signals from these layers
\citep{ma01}.  We follow \citet{basu04} in choosing the following
functional form:
\begin{eqnarray}
\delta^2 \nu & = & a_0 
+ b_0 \sin (4\pi\nu\taubcz\!+\!\phibcz) \nonumber \\
& & + (c_0\!+\!c_2/\nu^2) \sin(4\pi\nu\tauhiz\!+\!\phihiz),
\label{eq:abm08}
\end{eqnarray}
where $a_0$, $b_0$, $c_0$, $c_2$, \taubcz, \phibcz, \tauhiz\ and
\phihiz\ are 8 free parameters of fitting.  Since the number of data
points in the observed set of frequencies for \hdf\ is quite small, we
optimise the number of free parameters to strike a balance between a
fair representation of the oscillatory signal and a reasonable \chisq.
We find that the smooth component is fairly constant over the range of
frequencies that we use, as is the amplitude of the signal from the BCZ.
However, the amplitude of the signal from the HIZ varies more sharply
with frequency, and thus requires at least one frequency-dependent term.
We note that \citet{basu04} have shown that the exact form of the
amplitudes of the oscillatory signal does not affect the results
significantly.  To test our results, we also fit a function suggested by
\citet{hougou07} (simplified version of Eq.~(22) in that paper, with 8
free parameters).

The fit is carried out through a nonlinear \chisq\ minimisation,
weighted by the errors in the data. The errors in the second differences
are correlated, and this is taken care of by defining the \chisq\ using
a covariance matrix.  The effects of the errors are considered by
producing 1000 realisations of the data, where the mean values of the
frequencies are skewed by random errors corresponding to a normal
distribution.  The successful convergence of such a non-linear fitting
procedure is somewhat dependent on the choice of reasonable initial
guesses. To remove the effect of initial guesses affecting the final
fitted parameters, we carry out the fit for multiple combinations of
starting values.  For each realisation, $100$ random combinations of
initial guesses are tried for fitting the function above and from among
the resulting fits, the one which produces the minimum value of \chisq\
is accepted as the fit for that particular realisation.

The median value of each parameter for 1000 realisations is taken as its
fitted value. The error in the parameter is estimated from the range of
values covering $68\%$ area about the median (corresponding to $1\sigma$
error). As an example, the histogram of distribution of the parameters
\taubcz\ and \tauhiz\ over multiple realisations are illustrated in
Fig.~\ref{fig:histo_diffl}. Thus the quoted errors in these parameters
reflect the width of these histograms on two sides of the median value.

The strength of the oscillatory signals are measured through the
amplitudes of the respective oscillatory components in the frequencies
themselves. These are denoted by \abcz\ and \ahiz\ respectively, and
estimated by averaging the amplitudes of the oscillatory components in
Eq.~\ref{eq:abm08} over the observed frequency range, and correcting for
the scaling factor of $4\sin^2(\pi \tau/\tau_0)$ arising in the second
differences \citep[see][]{ma01}. In the present study, we use the
amplitudes only as an index of the significance of the detected depth of
the acoustic glitches.

We tested the technique on the frequencies of a polytropic model, with a
polytropic index of $1.5$, provided by \citet{jcd94}. Since there are no
sharp acoustic features in such a model, we do not expect to find any
oscillatory pattern in the second differences of the frequencies.
Indeed, if we attempt to fit a function of the form given by
Eq.~\ref{eq:abm08} to the second differences of this model, we only find
a signal corresponding to half of the large separation, which is the
periodicity at which the successive points occur due to $\ell=0$ or $2$
and $\ell=1$. 

The range of acoustic depth that can be probed by this technique is
limited. The smallest depth that can be fitted corresponds to the range
in the observed frequencies, and is $\sim 350$\,s. On the other hand,
the sampling frequency of the data points determines the largest
acoustic depth that can be fitted. The average large separation of \hdf\
being 85\,\muhz, the upper limit on the acoustic depth that can be
fitted is $\sim 2940$\,s.

\section{Results}
\label{sec:results}

\subsection{Fitting the CoRoT frequencies}
\label{subsec:results_data}

We use the second differences of the frequencies obtained by
\citet{benomar09} to extract the acoustic depths of BCZ and HIZ of \hdf.
We ignore frequencies which have errors of more than $3\,\muhz$. This
gives a total of 31 second differences to be used for $\ell=0,1,2$.  We
consider four different subsets of the data, namely: (i) $\ell=0$ modes
only (12 data points), (ii) $\ell=1$ modes only (16 data points), (iii)
$\ell=0,1$ modes (28 data points), and (iv) $\ell=0,1,2$ modes (31 data
points). This was done to test for consistency of the results when modes
of different degrees are considered, especially since the $\ell=0$ and
$\ell=2$ modes in the data have, on the average, higher errors than the
$\ell=1$ modes.  However, the number of data points in the first two
cases are only marginally larger than the number of free parameters in
the fitting function, and these cases are used only for testing
purposes.

The fits to the second differences of the mean frequencies and the
distribution of the fitted acoustic depths of BCZ and HIZ for different
realisations of the data are shown in Fig~\ref{fig:histo_diffl}.
Although we fit 1000 realisations of the data, only those realisations
with $\taubcz \leq 2700$\,s are considered due to reasons explained
below.  A fit to the data of subset (iv) above to the form suggested by
\citet{hougou07} is shown also in Fig~\ref{fig:histo_diffl}.
Table~\ref{tab:results} lists the median values of the acoustic depths
with estimated errors and the corresponding amplitudes of BCZ and HIZ
signals with respective estimated errors for the four data subsets.

\myfigure{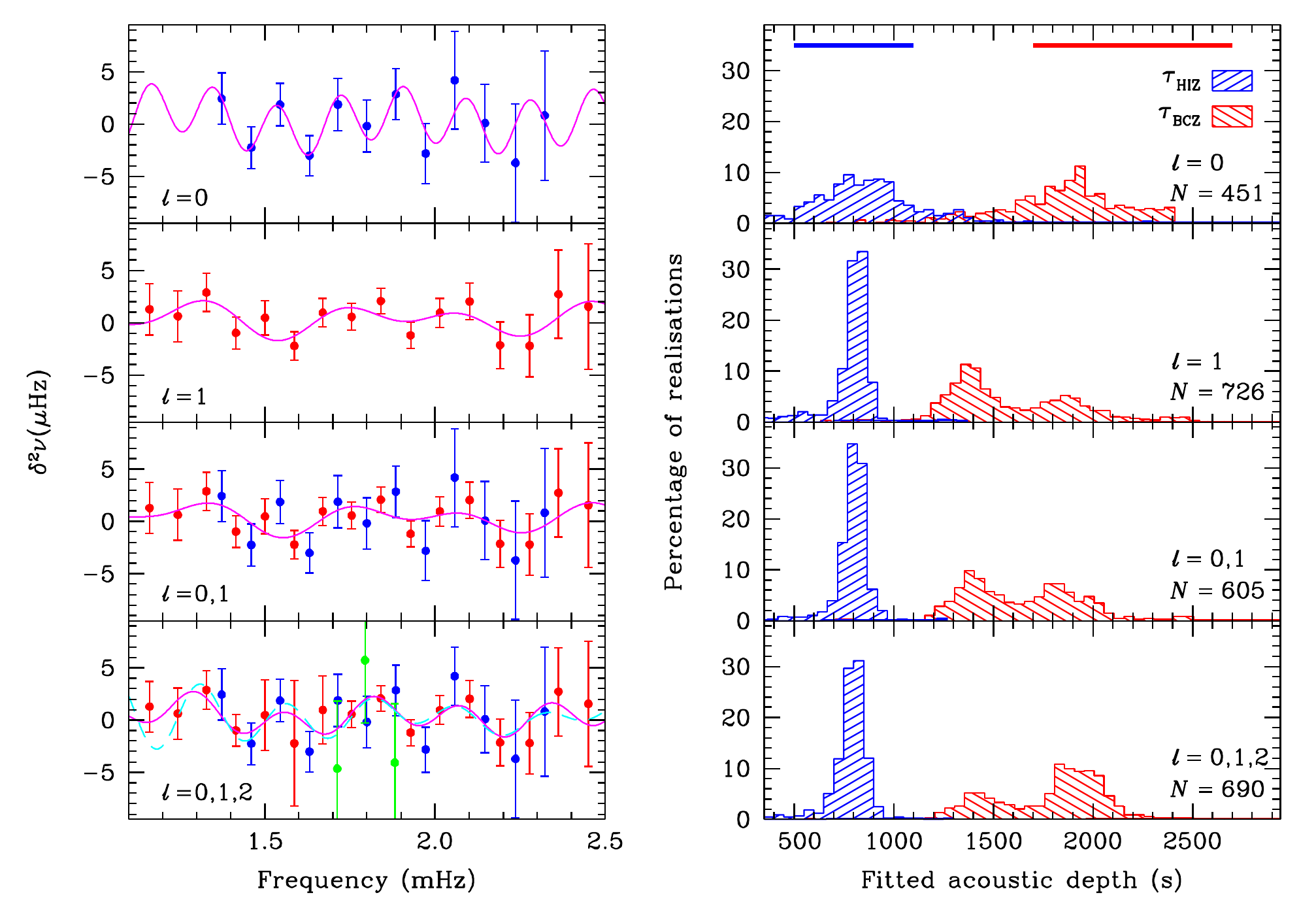}
{
Fits of Eq.~\ref{eq:abm08} to the second differences of the mean CoRoT
frequencies for \hdf\ and the histograms for the fitted values of
\tauhiz\ and \taubcz\ for different realisations of the data are shown
for different subsets (corresponding to different choices of $\ell$
values).  The second differences of the frequencies of $\ell=0$ ({\it
blue}), $\ell=1$ ({\it red}) and $\ell=2$ ({\it green}) modes of \hdf\
and their fit to Eq.~\ref{eq:abm08} ({\it magenta curve}) are shown in
the {\it left panel}.  The {\it dotted cyan} line in the {\it bottom
left panel} shows the fit by the \citet{hougou07} form.  The
corresponding histograms of the fitted values of \taubcz\ (in {\it red})
and \tauhiz\ (in {\it blue}) for different realisations are shown in the
{\it right panel}.  $N$ denotes the total number of valid realisations
after rejecting fits with $\taubcz > 2700$\,s.  The solid bands at the
top of the {\em right panel} indicate the range of initial guesses for
the two parameters in each fit.
}
{fig:histo_diffl}
{*}

\begin{table}
\setlength{\extrarowheight}{2.6pt}
\caption{Results of fitting second differences of \hdf\ frequencies to
Eq.~\ref{eq:abm08}. Median values of the acoustic depths and amplitudes
and their errors are listed.  The last two rows give the calculated and
the fitted values of a representative stellar model of \hdf.
\label{tab:results}
}
\begin{center}
\begin{tabular}{lcccc}
\hline
$\ell$ & \taubcz\ (s) & \tauhiz\ (s) & \abcz\ (\muhz) & \ahiz\ (\muhz) \\
\hline
0
      & ${1877}_{- 340}^{+ 273}$
      & ${ 847}_{- 253}^{+ 248}$
      & ${1.06}_{-0.34}^{+0.59}$
      & ${1.92}_{-0.70}^{+5.80}$
\\
\hline
1
      & ${1496}_{- 194}^{+ 438}$
      & ${ 812}_{-  79}^{+  45}$
      & ${0.75}_{-0.20}^{+0.24}$
      & ${1.47}_{-0.29}^{+0.33}$
\\
\hline
0,1
      & ${1626}_{- 262}^{+ 327}$
      & ${ 806}_{-  61}^{+  45}$
      & ${0.66}_{-0.15}^{+0.18}$
      & ${1.40}_{-0.22}^{+0.28}$
\\
\hline
0,1,2
      & ${1855}_{- 412}^{+ 173}$
      & ${ 794}_{-  68}^{+  55}$
      & ${0.71}_{-0.15}^{+0.19}$
      & ${1.32}_{-0.31}^{+0.34}$
\\
\hline
Model
      & $1998$
      & $882$
      &
      &
\\
0,1,2
      & ${1982}_{- 586}^{+ 447}$
      & ${ 848}_{-  62}^{+  61}$
      & ${0.47}_{-0.13}^{+0.19}$
      & ${1.37}_{-0.22}^{+0.24}$
\\
\hline
\end{tabular}
\end{center}
\end{table}

The acoustic depth of the HIZ obtained from the different sets are
consistent within the errorbars. The amplitude of the HIZ oscillatory
function is also significant at more than $2.5\sigma$ level in each set.

However, the BCZ oscillatory signal is not so consistent among different
sets. For all values of $\ell$, we get a peak in the histogram close to
the value $\tauzero/2 \sim 2900$\,s, which corresponds to the Nyquist
frequency. The occurrence of this peak could mean that the amplitude of
the oscillatory function corresponding to BCZ is close to zero, so that
the fitted sine wave has periodicity equal to the natural spacing of the
data points, {\em i.e.}, the large separation. While this seems to be
the case for $\ell=1$ modes, the second differences of the radial modes
appear to have an actual oscillatory signal corresponding to this
frequency! Thus this effect cannot be fully ascribed to the Nyquist
effect. This is also borne out by the fact that this peak persists even
when we combine data with multiple values of $\ell$, which should not be
the case \citep[see, e.g.,][]{ma01}. We cannot find a satisfactory
explanation as to the origin of this periodicity in the radial modes,
apart from an artifact introduced due to the noise.  In order to
eliminate this artifact for all subsets, we discard the realisations for
which the BCZ signal is fitted with an acoustic depth of around 2900\,s.
Few realisations yield \taubcz\ values higher than the acoustic midpoint
of the star ($\sim 2900$\,s), and these are discarded also.  Actually,
we consider only those realisations for which $\taubcz < 2700$\,s, since
the peak in the histogram around the acoustic midpoint has a typical
spread of 200\,s on either side of the peak. This results in reduction
of the number of valid realisations by nearly 30\%. 

In all the cases, there is a peak in the histogram for \taubcz\ around
1950\,s, which is most likely due to the base of the convective envelope
in the star. For $\ell=0$ and $\ell=0,1,2$, this is the dominant peak of
the histogram, while for $\ell=1$ and $\ell=0,1$, this is the secondary
peak. For the latter two cases, the dominant peak is centred around
1370\,s. If this value is to correspond to \taubcz\, it would imply a
very shallow convection zone in the star, which is an unlikely scenario
given the expected mass and age of \hdf. This peak cannot be associated
with any artifact due to the sampling frequency, but on the other hand,
we cannot assign any physical significance to it.

To understand the difference in the values of \taubcz\ obtained for
different $\ell$, we carried out the following test. For each case, we
held the parameter \taubcz\ in Eq.~\ref{eq:abm08} fixed at a certain
value, while allowing all other parameters to be free, and monitored the
value of the reduced \chisq\ obtained by the resulting fit to the mean
second differences. The fixed value of \taubcz\ was changed continuously
between 1000\,s and 4800\,s to find the minima in the reduced \chisq\ as
this parameter varies. The \chisq\ is plotted as a function of \taubcz\
in Fig.~\ref{fig:fixed_taub_chisq}. We find that the $\ell=1$ modes show
a deep minimum around 1370\,s, but only a slight inflexion around
1950\,s. In contrast, the location of the minimum and the inflexion are
interchanged for the $\ell=0$ modes. When these modes are combined along
with the $\ell=2$ modes, we have minima at both the values.  The third
minimum around 2900\,s for all cases has already been discussed.  The
almost mirrored shapes of the graphs above \taubcz\ values of 2900\,s
merely illustrate the aliasing effect, as elaborated in \citet{ma01}.

\myfigure{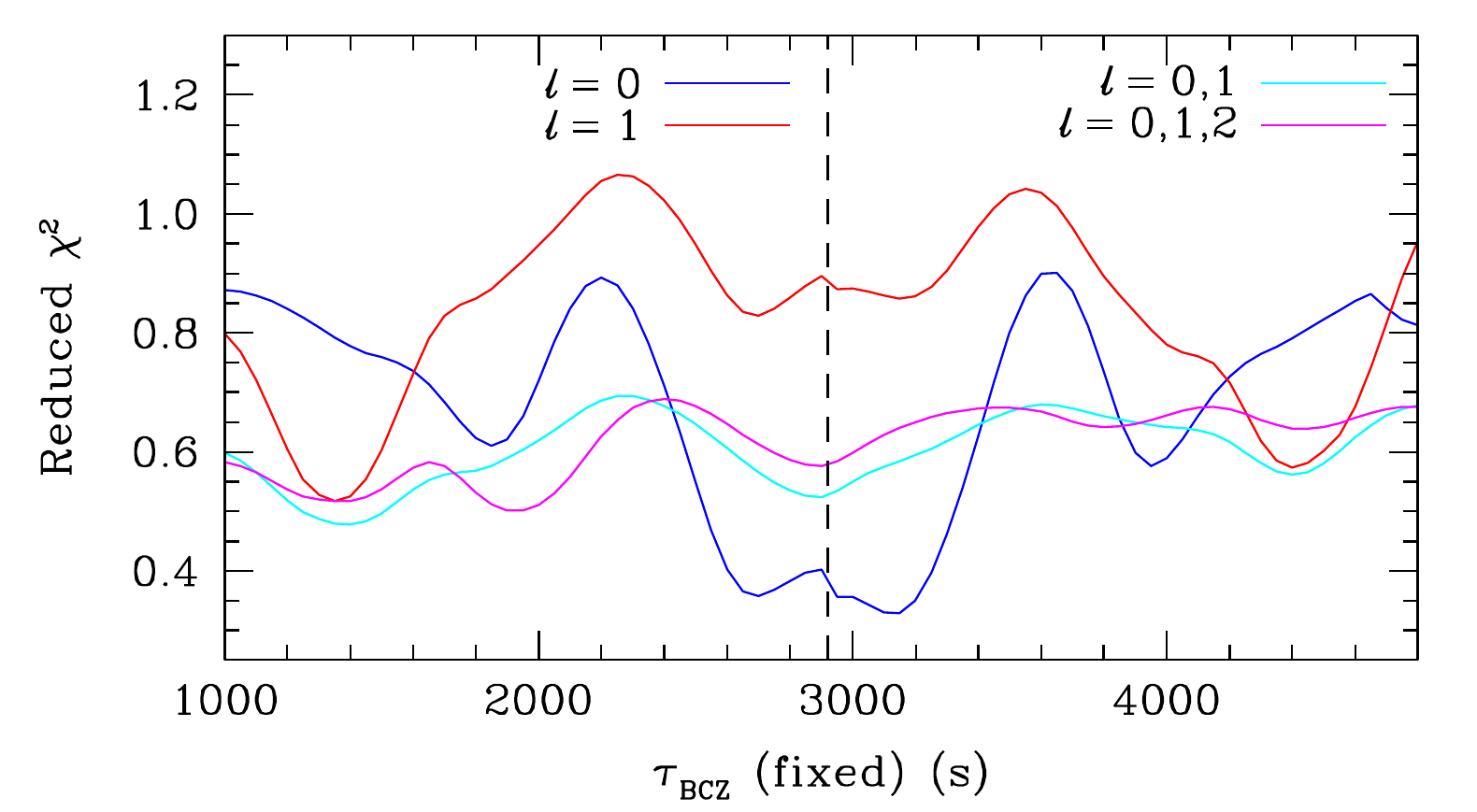}
{The variation of \chisq\ with the adopted value of \taubcz\ is shown
for four subsets of the CoRoT data on \hdf\ corresponding to different
choices of the degree, $\ell$. The vertical dotted line shows the
acoustic mid-point of the star.}
{fig:fixed_taub_chisq}
{}

\subsection{Fitting model frequencies}
\label{subsec:results_model}

We fit Eq.~\ref{eq:abm08} to the second differences of frequencies
calculated from a representative stellar model for the star \hdf\ as
well. The model is generated with the CESAM2k evolution code
\citep{ml08} and uses standard physical inputs, including convective
core overshoot, but no diffusive mixing.  While we use the theoretical
frequencies as the mean values in this case, we set the errors in the
frequencies equal to the errors in the observed frequencies for the
corresponding mode, in order to mimic the uncertainties present in the
data. The results of such a fit are displayed in
Fig.~\ref{fig:hd49933_model}.  The model values as well as the fitted
median values of \tauhiz\ and \taubcz\ are given in
Table~\ref{tab:results}.
\myfigure{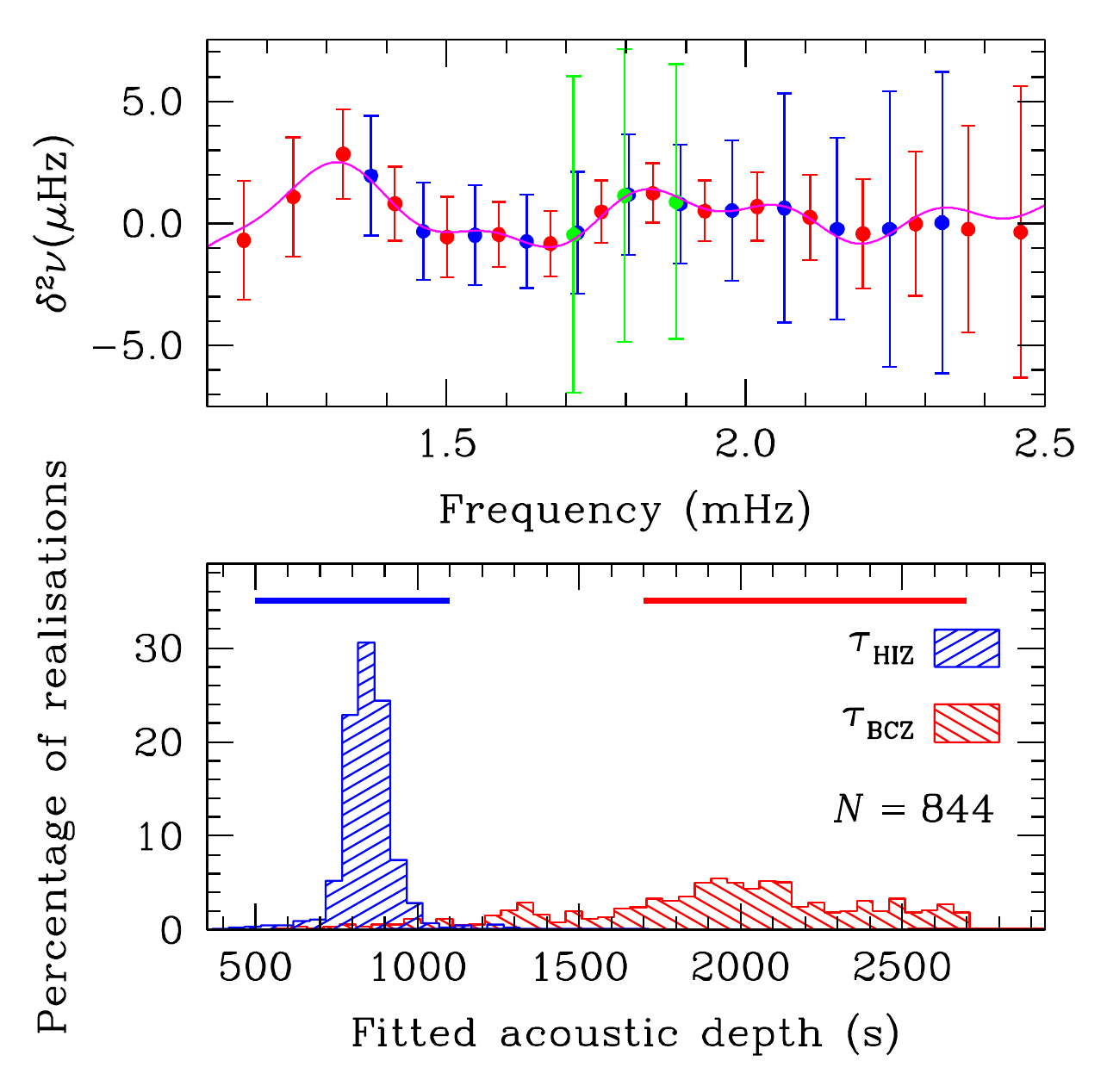}
{
Fits of Eq.~\ref{eq:abm08} to the second differences of the frequencies
of a theoretical model for \hdf\ ({\em top panel}) and the histograms
for different realisations of the data for the fitted values of \tauhiz\
and \taubcz\ ({\em bottom panel}) are shown.  See
Fig.~\ref{fig:histo_diffl} for explanations of colours and symbols.
}
{fig:hd49933_model}
{}

We find that the median values of the acoustic depths of both BCZ and
HIZ are recovered from the frequencies. However, the error in the value
of \taubcz\ is quite large due to the flat nature of the histogram in
Fig.~\ref{fig:hd49933_model}. In contrast, the histogram of the HIZ
signal is sharply peaked and hence the estimated error in the median
value of \tauhiz\ is also smaller. We conclude that the intrinsic
oscillatory signal arising from the acoustic glitch at BCZ is quite
weak, and it is difficult to extract this signal even with model
frequencies. It is beyond the scope of this study to investigate why
this signal is so weak, and it needs to be probed whether this is a
common feature of all stars in the relevant part of the \hrd.  The HIZ
signal, however, is strong and can easily be used to extract the
acoustic depth of the second helium ionisation zone.

\section{Summary}
\label{sec:summary}

We have applied the method of determining the acoustic depth of layers
of sharp variation in sound speed in stellar interiors from the
oscillatory signal in frequencies to the CoRoT primary target \hdf. 

The oscillatory signal is made more pronounced in comparison to the
smooth change in sound speed by using the second differences of the
frequencies. We fit a function with two oscillatory components
corresponding to the two major layers of sharp variation in sound speed,
{\em viz.}, the base of the convective envelope and the second helium
ionisation zone.  To prove that the technique does not produce
artificial signatures of oscillatory signals when there are none, we
tested it on the frequencies of a polytropic model where there are no
real layers of sharp variation in the sound speed. In this case the only
significant oscillatory signal corresponds to the uniform spacing of the
frequencies, as expected.

We find that the oscillatory signal arising due to the second helium
ionisation zone is quite strong in the CoRoT data of \hdf. We estimate
the acoustic depth of this layer to be $794_{-68}^{+55}$\,s. This is
borne out by the frequencies of modes of all available degrees:
$\ell=0,1,2$, when considered either separately, or in combination. The
acoustic depth of this layer in a representative stellar model is
consistent with this value, and can also be easily extracted by our
technique.

The oscillatory signal due to the base of the convective envelope is far
weaker in the data, in comparison to the helium signal. Moreover, we
find that the signal is not consistent between modes of different
degrees. We find additional periodicities in the frequencies
corresponding to different acoustic depths in modes of $\ell=0$ and
$\ell=1$, for which we are unable to assign any physical acoustic
feature inside the star. It seems likely that these are artifacts due to
the errors in the data which have no physical relevance. Nevertheless,
we do find an oscillatory signal corresponding to an acoustic depth of
$1855_{-412}^{+173}$\,s in the combined frequency set of different
degrees, which is consistent with the position of the base of the
convective envelope in a representative stellar model. This oscillatory
signal is weak even in the model frequencies, and is fairly difficult to
extract from the second differences.  These results prove that the
technique of extracting acoustic depths of sharp features inside a star
can be successfully applied for real asteroseismic data. 

\begin{acknowledgements}
CoRoT (Convection, Rotation and planetary Transits) space mission has
been developed and is operated by the French Space agency CNES in
collaboration with Austria, Belgium, Brazil, ESA's  RSSd and Science
Programmes, Germany and Spain.  AM acknowledges support from the
National Initiative on Undergraduate Science (NIUS) undertaken by the
Homi Bhabha Centre for Science Education -- Tata Institute of
Fundamental Research (HBCSE-TIFR), Mumbai, India.
\end{acknowledgements}


\begin{thebibliography}{}

\bibitem[Baglin et al.(2006)]{baglin06} 
Baglin, A., Auvergne, M., Barge, P., et al.\ 
2006, ESA SP, 1306, 33 

\bibitem[Ballot et al.(2004)]{btg04} 
Ballot, J., Turck-Chi{\`e}ze, S., \& Garc{\'{\i}}a, R.~A.\ 
2004, \aap, 423, 1051

\bibitem[Basu \ea(1994)Basu, Antia, \& Narasimha]{ban94}
Basu, S., Antia, H.~M., \& Narasimha, D.\
1994, \mnras, 267, 209

\bibitem[Basu et al.(2004)]{basu04} 
Basu, S., Mazumdar, A., Antia, H.~M., \& Demarque, P.\ 
2004, \mnras, 350, 277 

\bibitem[Benomar et al.(2009)]{benomar09}
Benomar, O., Baudin, F., Campante, T.~L., et al.\ 
2009, \aap, 507, L13 

\bibitem[Christensen-Dalsgaard \& Mullan(1994)]{jcd94}
Christensen-Dalsgaard, J., \& Mullan, D.~J.\ 
1994, \mnras, 270, 921 

\bibitem[Gough(1990)]{gough90} 
Gough, D.~O.\ 
1990, Progress of Seismology of the Sun and Stars, 367, 283 

\bibitem[Houdek \& Gough(2007)]{hougou07}
Houdek, G., Gough, D.~O.\ 
2007, MNRAS, 375, 861

\bibitem[Mazumdar(2005)]{am05} 
Mazumdar, A.\ 
2005, \aap, 441, 1079 

\bibitem[Mazumdar \& Antia(2001)]{ma01} 
Mazumdar, A., \& Antia, H.~M.\ 
2001, \aap, 377, 192 

\bibitem[Miglio et al.(2010)]{miglio10} 
Miglio, A., Montalb{\'a}n, J., Carrier, F., et al.\ 
2010, \aap, 520, L6 

\bibitem[Monteiro \ea(1994)Monteiro, Christensen-Dalsgaard, \& Thompson]{mct94}
Monteiro, M.~J.~P.~F.~G., Christensen-Dalsgaard, J., \& Thompson, M.~J.\
1994, \aap, 283, 247

\bibitem[Monteiro et al.(2000)Monteiro, Christensen-Dalsgaard, \& Thompson]{mct00} 
Monteiro, M.~J.~P.~F.~G., Christensen-Dalsgaard, J., \& Thompson, M.~J.\ 
2000, \mnras, 316, 165

\bibitem[Morel \& Lebreton(2008)]{ml08}
Morel, P., \& Lebreton, Y.\
2008, \apss, 316, 61

\bibitem[Roxburgh \& Vorontsov(1994)]{rv94}
Roxburgh, I.~W.~\& Vorontsov, S.~V.\
1994, \mnras, 268, 880

\bibitem[Roxburgh \& Vorontsov(2003)]{rv03}
Roxburgh, I.~W.~\& Vorontsov, S.~V.\
2003, \aap, 411, 215

\end{thebibliography}
\end{document}